\documentclass[journal]{IEEEtran}
\usepackage{graphicx}
\usepackage{hyperref}
\usepackage{url}
\usepackage{soul}
\usepackage{xcolor}

\usepackage{cite}

\ifCLASSINFOpdf
\else
\fi

\begin{document}
%





\title{Bridging the Urban-Rural Connectivity Gap through Intelligent Space, Air, and Ground Networks}


%
%
%

\author{
        Fares~Fourati,
        Saeed~Hamood~Alsamhi, and Mohamed-Slim~Alouini
\thanks{Fares Fourati and Mohamed-Slim Alouini are with King Abdullah University of Science and Technology (KAUST), Thuwal, 23955-6900 Kingdom of Saudi Arabia, (e-mail: fares.fourati@kaust.edu.sa, slim.alouini@kaust.edu.sa). S.H.  Alsamhi is with the Athlone  Institute of Technology, Ireland,  and Ibb University, Ibb, Yemen. (e-mail: salsamhi@ait.ie).}
}

%
%

\markboth{2021}
{\MakeLowercase{\textit{et al.}}: }
%



\maketitle 

\begin{abstract}

Connectivity in rural areas is one of the main challenges of communication networks. To overcome this challenge, a variety of solutions for different situations are required. Optimizing the current networking paradigms is therefore mandatory. The high costs of infrastructure and the low revenue of cell sites in rural areas compared with urban areas are especially unattractive for telecommunication operators. Therefore, space, air, and ground networks should all be optimized for achieving connectivity in rural areas. We highlight the latest works on rural connectivity, discuss the solutions for terrestrial networks, and study the potential benefits of nonterrestrial networks. Furthermore, we present an overview of artificial intelligence (AI) techniques for improving space, air, and ground networks, hence improving connectivity in rural areas.  AI enables intelligent communications and can integrate space, air, and ground networks for rural connectivity. We discuss the rural connectivity challenges and highlight the latest projects and research and the empowerment of networks using AI. Finally, we discuss the potential positive impacts of providing connectivity to rural communities.
\end{abstract}


%
\IEEEpeerreviewmaketitle

\newcommand{\RNum}[1]{\uppercase\expandafter{\romannumeral #1\relax}}

\section{Introduction} 

\begin{figure}
    \centering
    \includegraphics[scale = 0.44]{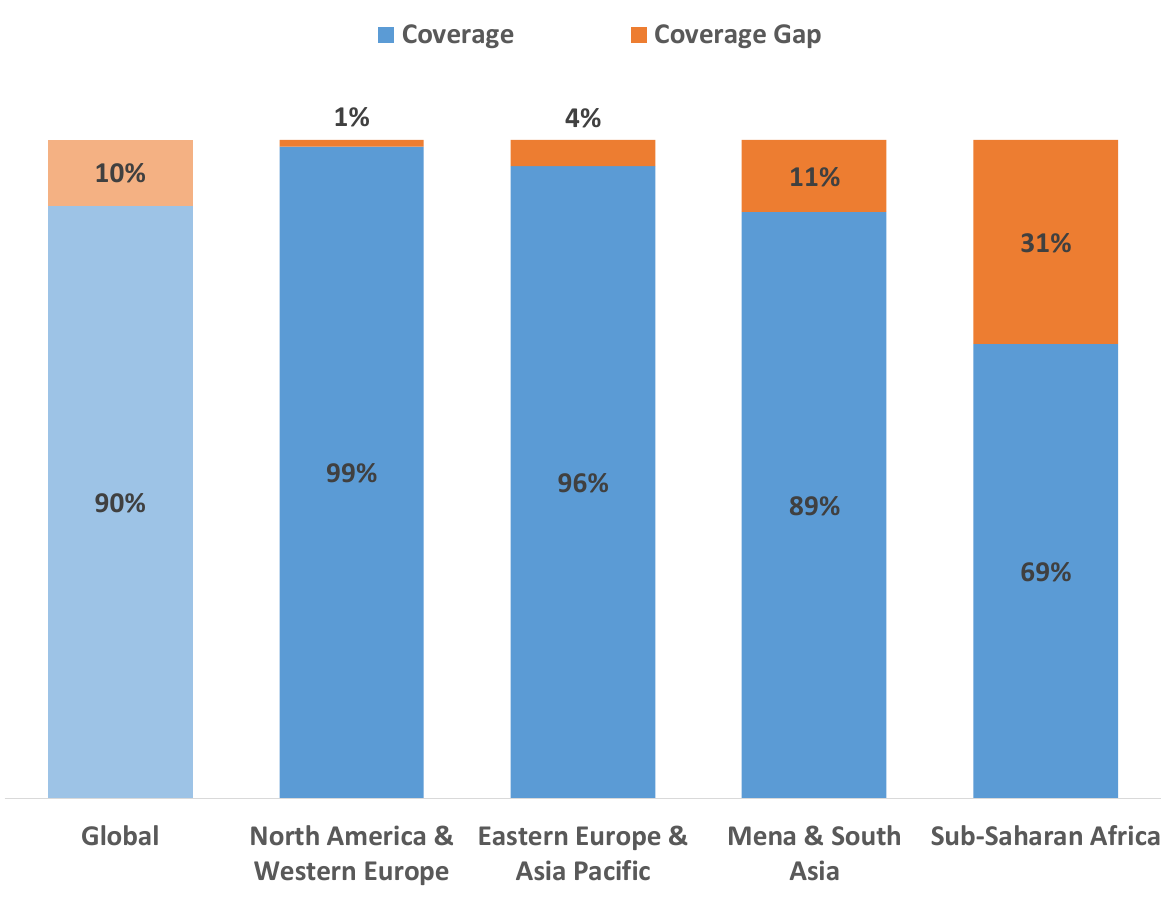}
    \caption{Coverage gaps in areas around the world \cite{bahia2019state}.}
    \label{gsma}
\end{figure}

\IEEEPARstart{I}{n} most of the remote, hard-to-reach, and rural areas worldwide, internet connectivity is possible only via mobile access \cite{kusuma2021diffractive}. Fortunately, a remarkable progress in providing worldwide internet connectivity has been made by the mobile industry \cite{gsma}. For example, in 2018, 300 million new mobile internet subscribers joined the network \cite{bahia2019state}. Nevertheless, over 750 million people still lack mobile broadband coverage \cite{bahia2019state}. Coverage depends on various features such as demography, geography, and economy. Consequently, this lack of coverage is not uniformly distributed and is accentuated in remote and rural areas, especially in areas such as subSaharan Africa, where 31\% of the population is deprived of access to connectivity (Fig. \ref{gsma}) \cite{bahia2019state}.

Although the main purpose of connectivity is facilitating the communication between people, connectivity can provide a wide spectrum of applications that improve the quality of life of users by providing useful information and services as well as opportunities for enhancing economic growth \cite{katz2018economic}. If provided with a high-quality connection, people in rural areas can remain in their quiet villages while enjoying the advantages of the city, working from home, remote e-learning on advanced platforms, and other activities \cite{9301389}.

Besides connecting people, connectivity in rural areas can be extended to introduce smart frameworks by connecting devices, collecting data, preforming real time processing and monitoring, and using artificial intelligence (AI) in different scenarios. For instance, in agriculture, connectivity can improve the efficiency of production processes by connecting sensors, collecting different types of data, and using algorithms to help with decision making and early diagnosis of problems. Moreover, in deep rural areas, the Sahara, mountains, and forests, connectivity allows exploration and monitoring for security or scientific research purposes. To achieve rural connectivity and benefit from its advantages, several challenges must be overcome.

\subsection{Main Challenges}

In any business, an investment should offer a return. In rural and remote areas, however, the revenue is lower than that in urban areas; therefore, infrastructure costs can be prohibitive. In addition, the logistics are generally complex, especially for hard-to-reach areas. In fact, the revenue can be 10-times higher in urban sites than that in rural areas and the network infrastructure cost in urban areas is generally around half of that in rural areas \cite{gsma}.

The Global System for Mobile Communications Association (GSMA) has identified three main applications in which the infrastructure costs can be restrictive: mobile base stations (BSs), which supply coverage to an area; backhaul technology, which routes user-generated voice calls and data to the core network, including wireless and wired technologies; and the energy that enables both of these components to function \cite{gsma}. Cabling is costly and long-distance data transmission through wireless technology requires the use of several receivers and transmitters that are also costly \cite{gsma}. Although mobile operators heavily invested in the deployment optimization of their networks, connecting the unconnected areas remains a major problem. 


 

\subsection{Related Surveys and Contributions}
A detailed survey on rural connectivity has been published \cite{9042251}. Fronthaul and backhaul solutions for Internet of Things (IoT) in rural areas are discussed in \cite{9301389}. The use of AI for the optimization of communication systems has been reviewed in various papers, including the use of AI in satellite communications \cite{fourati2021artificial}, and unmanned aerial vehicules (UAV) communications \cite{lahmeri2021artificial}.

As the interest in rural connectivity continues to evolve, several solutions, ideas, and papers have been published lately discussing this subject. In this study, we mention the latest works on rural connectivity, discuss the solutions for terrestrial networks, and study the potential applicability of nonterrestrial networks. 
Technologies for rural connectivity have been thoroughly discussed elsewhere \cite{9042251,9301389}; thereofre, they are only brifely overviewed here with focus on the latest works. Although AI-based optimization of communication systems has been reported in previous studies \cite{fourati2021artificial, lahmeri2021artificial}, we discuss the role of AI in empowering networks in rural areas. This discussion will emphasize the role of AI in future communication systems, especially in challenging contexts such as rural areas. Finally, we discuss the potential positive impacts of providing connectivity to rural communities.

In this study, we view rural connectivity from different perspectives, beginning with the latest works on terrestrial, aerial, and space networks, progressing to the discussion of the potential role of AI in efficiently empowering and optimizing these networks in rural areas, and concluding with the potential impacts of rural connectivity on rural communities.

\subsection{Paper Organization} 

The remained of this paper is organized as follows. Section \ref{2} discusses the use of latest terrestrial, aerial, and space technologies for rural connectivity and concludes with the combination of these networks for connectivity. Section \ref{3} refers to the potential applications of AI in different communication sectors. In this section, we discuss the further optimization of each communication type and their intersection. Section \ref{4} highlights the potential benefits of rural connectivity and the use of AI in different sectors, including health, education, agriculture and economy sectors of rural areas.

\section{Space, Air, and Ground Networks}\label{2}

\subsection{Use of Terrestrial Technologies}

The standard method of providing cellular coverage involves the use of macrocell sites. In challenging rural areas, several macrocell sites must be built to provide coverage, leading to a poor cost-to-coverage ratio. Meanwhile, the capacity must sufficiently cover each user. To provide high capacity under the cost-to-coverage constraints, Meta/Facebook proposed a novel connectivity solution called SuperCell, which has been field-validated for feasibility. SuperCell includes two essential components: a high-altitude tower with high-gain antennas for wide coverage and high-order sectorization to offer high capacity through heavy frequency reuse. A single SuperCell procides 15-times to 65-times wider coverage than macrocells. More specifically, the coverage area of a 36-sector SuperCell BS mounted on a 250-m-high tower is up to 65-times that of a standard 3-sector rural macrocell BS on a 30-m-high tower in an identical arrangement, implying that the single SuperCell can replace multiple macrocells \cite{fb}. After some engineering investment, commercially-available antenna systems may enable a highly efficient SuperCell for wide-area rural connectivity \cite{bondalapati2020supercell}.

Terrestrial microwave backhaul is primarily based on clear line-of-sight (CLOS) design. When a CLOS is unavailbe, network designers usually implement repeater design topologies or intermediate site passive repeaters, which lead to prohibitive costs. An altrenative solution nonline-of-sight (NLOS) links, which exploit the natural diffraction propagation effects that scatter the radio signal around a blocking obstacle. The main challenges in this technique are accurately forecasting of the operation and performance of NLOS radio links at high frequencies (6-8 GHz) and forecasting the availability performance of the NLOS radio links over time. After investigating various design strategies, Meta/Facebook proposed a hybrid design approach by combining NLOS and CLOS \cite{kusuma2021diffractive}. When both links are used together, the number of repeaters can be notably decreased, enabling more flexible choices of the tower positions and sizes, thereby improving coverage. This hybrid design can attain a similar coverage objectives with remarkably lower cost than using CLOS designs \cite{kusuma2021improving}. To prove this claim, Meta/Facebook collaborated with Mayutel and Internet para Todos of Peru to check this idea in a real world application, which was successfully implemented in Peru \cite{kusuma2021improving}.

\subsection{Use of Aerial Technologies}

Over the past few years, the communication research community has used the emergence of UAVs to suggest their integration into communication networks. UAV BSs are intrinsically, mobile, flexible, and height-adjustable \cite{8660516}. Therefore, they can complement current cellular systems by offering broadband coverage in hard-to-reach rural areas. UAVs give operators more flexibility regarding site placement. As BSs become more miniaturized and UAVs become cheaper and more refined, implementing BSs on UAVs has become feasible in practice.  As UAV BSs can be expeditiously fastened at ideal positions in three-dimensional space, they can provide higher spectral efficiency, coverage, load balancing, and user experience compared with those using current terrestrial technologies \cite{8675384}.

\begin{figure}
    \centering
    \includegraphics[scale = 0.56]{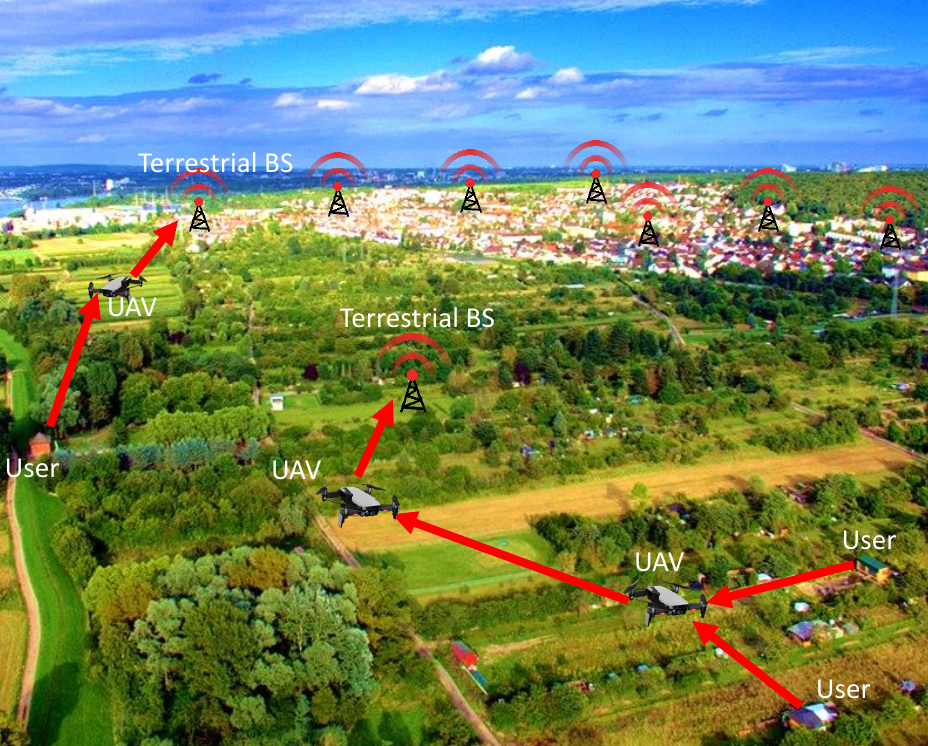}
    \caption{Combined aerial and terrestrial networks for remarkably enhancing the quality of service in rural and hard-to-reach areas \cite{9420290}.}
    \label{999}
\end{figure} 

With their competitively low cost, high mobility, and quick implementation UAVs become suitable and efficient for covering low-traffic large regions. Moreover, they are suitable for precision agriculture \cite{daponte2019review} and delivery \cite{7513397}.

In \cite{9420290}, the spatial variation of coverage among urban, suburban, and exurban areas was studied and a stochastic geometry-based model was designed using inhomogeneous Poisson point processes, studying a scenario where ground BSs were gathered in the urban area while aerial BSs were uniformly distributed outside of that urban area (see Fig. \ref{999}). The authors \cite{9420290} showed that combining the terrestrial network with aerial BSs in the outer regions generally enhances the quality of service.

In \cite{galan2021energy}, an optimization problem was designed for planning a set of assignments at different altitudes of a swarm of UAVs, offering rural area coverage while minimizing the energy consumption. The problem was solved using a genetic algorithm model and assessed over a realistic setting. Increasing the number of altitudes at which the UAVs can be positioned was found to increase coverage but require more energy than when all UAVs occupied one altitude.

A popular solution is the use of Altaeros SuperTower \cite{altaeros, belmekki2020unleashing}, an aerostatic helium-filled blimp tethered at 240 m that provides a wide area coverage of up to 10,000 km$^2$. Altaeros SuperTower provides the coverage equivalent to those using more than 20 towers, but at a lower cost and with a simpler installation process. Each tether of the aerostat includes fiber optics and a power conductor for providing electricity. The aerostat is 5G-ready but does not support 2G and 3G, which can be a disadvantage in poor rural regions \cite{altaeros,gsma}.

\subsection{Use of Space Technologies}

Like UAVs, satellites provide an interesting solution to overcome geographic constraints and to skip installation delays, especially in the areas with no existing infrastructure. Previously, geostationary earth orbit (GEO) satellites were commonly used for this purpose because they avoid the need for quick movements between the terminals and the satellite transceiver and permit for a wide coverage using a single satellite \cite{9210567}. 

More recently, new communication technologies and cheaper launch costs have inspired the development of satellite constellations such low earth orbit (LEO) satellite constellations, which provide low latency, high-throughput broadband services \cite{9210567}. 

\begin{figure}
    \centering
    \includegraphics[scale = 0.55]{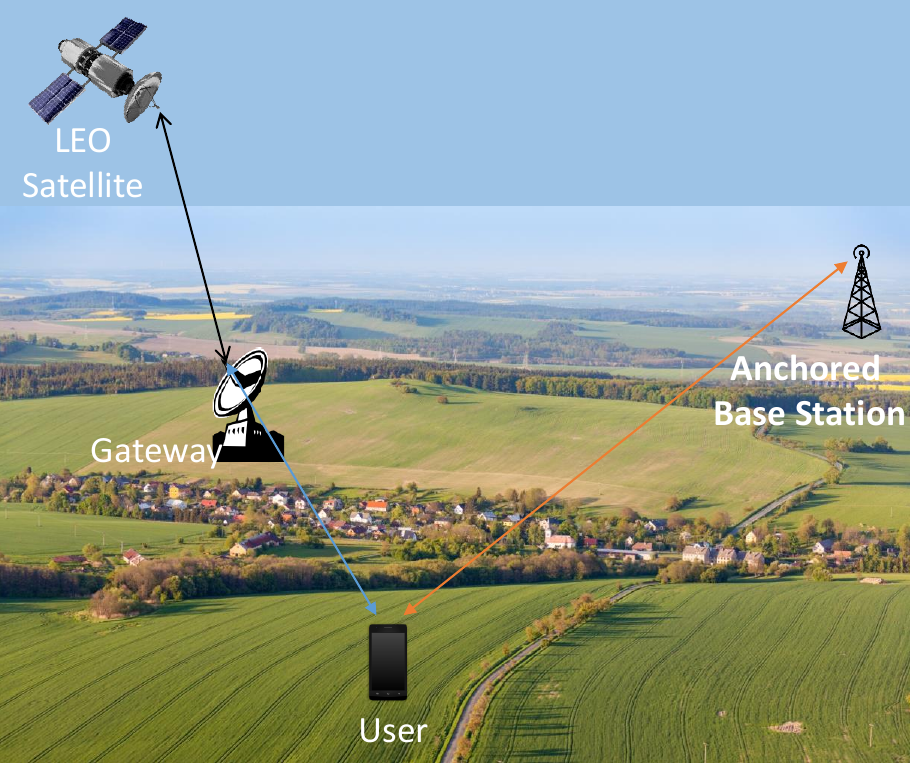}
    \caption{Satellite networks can remarkably improve coverage for rural and hard-to-reach areas  \cite{9218989}.}
    \label{sagin3}
\end{figure}

The author of \cite{9218989} studied the performance of LEO constellation, where the locations of the satellites were modeled as a binomial point process on a spherical surface (Fig. \ref{sagin3}). Comparing the system performances of the LEO constellation and a fiber-connected BS at a relatively far distance, they found that the LEO constellation improved the coverage probability of rural regions. 

As the locations of GEO satellite are fixed relative to the ground, they admit inexpensive user equipment. LEO and medium earth orbit (MEO) constellations reduce the latency but need more complex antennas at the satellite and ground stations for satellite tracking \cite{7565492}, \cite{9042251}. Several companies, such as Amazon, SpaceX, and OneWeb are developing LEO solutions \cite{9210567}, whereas MEO constellations such as that suggested by O3b \cite{o3b} are expected to balance the constellation latency-size tradeoff \cite{9210567}.


\subsection{Integration of Space-Air-Ground Technologies}

\begin{table}
\caption{Comparison of Different Technologies \cite{sagin2, sagin1}.}
\setlength{\tabcolsep}{5pt}
\begin{tabular}{|p{38pt}|p{91pt}|p{91pt}|}
\hline
Technology & Pros & Cons\\
\hline
\hline
 Ground  & Rich resources and rich throughput & Limited coverage, difficult movement, and vulnerable to disasters\\
\hline
 Aerial  & Wide coverage, low cost, and flexible deployment & Less capacity, unstable link, and high mobility\\
\hline
 Space  & Large coverage and broadcast/multicast & Long propagation latency, limited capacity, high cost, and high mobility\\
\hline
\end{tabular}
\label{tabb}
\end{table}

\begin{figure}
    \centering
    \includegraphics[scale = 0.4]{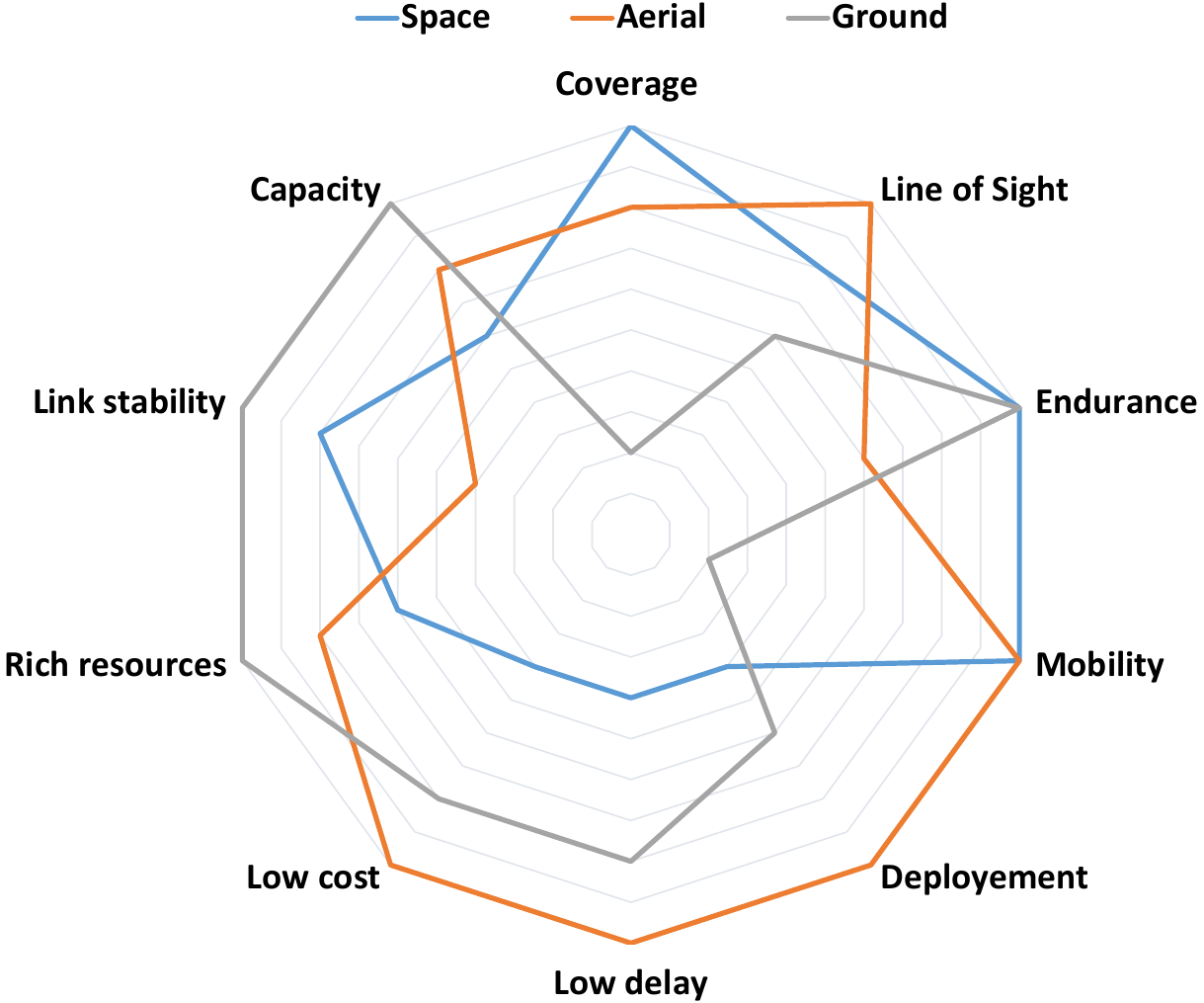}
    \caption{Comparison of space, air, and ground technologies  \cite{belmekki2020unleashing,kishk2020aerial,sagin1}.}
    \label{sagin}
\end{figure}

\begin{figure}
    \centering
    \includegraphics[scale =0.51]{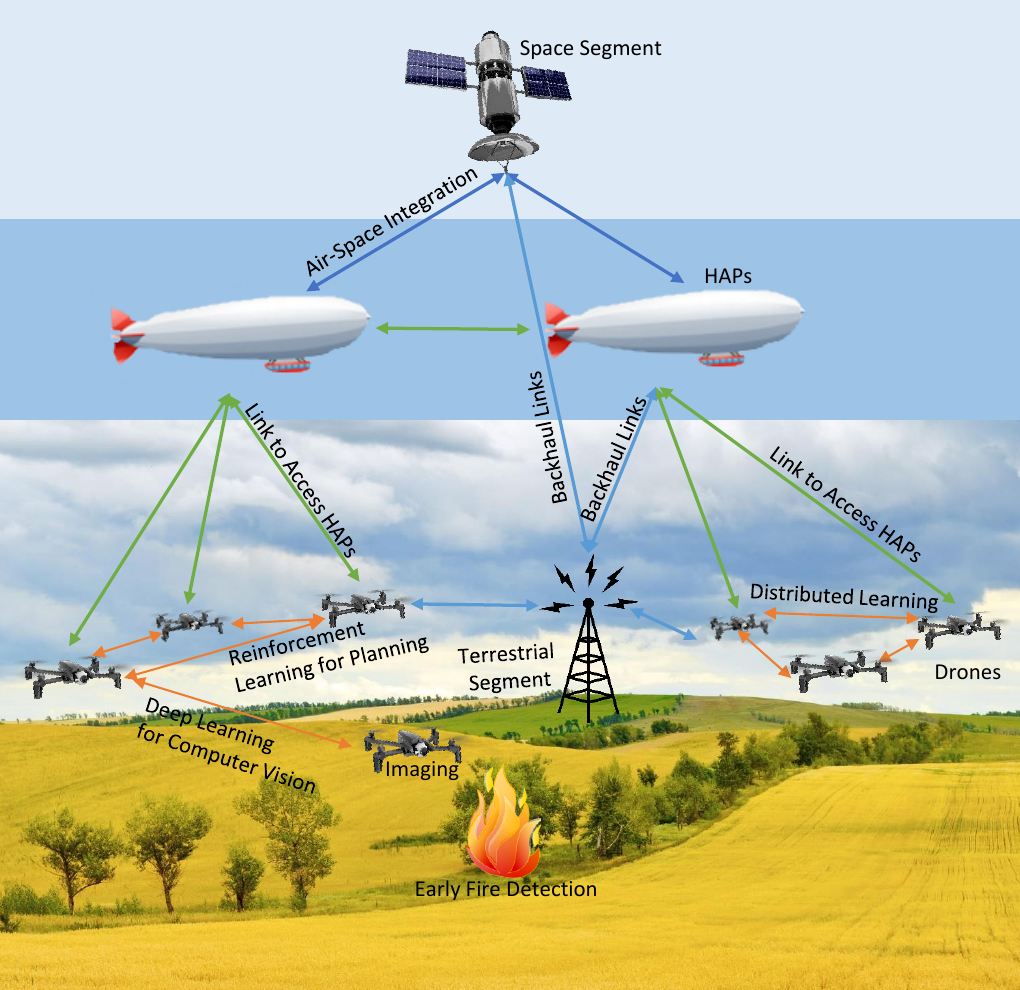}
    \caption{AI supporting the integration of space, air, and ground networks in rural areas areas.}
    \label{saginai}
\end{figure}

As summarized in Table \ref{tabb}, each of the discussed technologies has its advantages and limitations. Terrestrial networks have the lowest delay among the various communication technologies but are exposed to natural disasters. Air networks provide wide coverage with low latency but are limited by restricted capacity and unstable links. Satellite networks offer global coverage but suffer from high latency (Fig.\ref{sagin}) \cite{sagin2}, \cite{sagin1}. Recognizing that have different characteristics and can complement each other, researchers have proposed space, air, and ground integrated networks (SAGINs) for optimized end-to-end services \cite{sagin2}. SAGINs include the GEO, MEO, and LEO satellites in space; the balloons, airships, and drones in the air; and ground segments (see Fig. \ref{saginai}). The multilayered satellite communication system can inmprove the network capacity using the multicast and broadcast methods \cite{sagin2}.  However, SAGINs are inherently time-variable and heterogeneous, which complicates their design and optimization \cite{sagin2}. In decision making, it is necessary to consider the high mobility of the space and air segments, the diverse propagation mediums, the sharing of frequency bands, and the intrinsic heterogeneity between the space, air, and ground segments. Although, SAGIN optimization is challenging, efficient techniques for integration could open opportunities for rural connectivity.

\section{AI-empowered Networks} \label{3}


\subsection{AI for Space Networks}
AI, especially machine learning \cite{pattern}, has been widely and diversely applied in recent years. For the optimization of different communication aspects, several solutions have been suggested using AI algorithms \cite{zhu2020toward, sun2019application, morocho2019machine}. In fact, AI can be useful to optimize each segment. For example, in satellite communication (the space segment) \cite{fourati2021artificial}, AI has demonstrated promise in behavior modeling, channel modeling, network traffic forecasting, remote sensing, interference managing, ionospheric scintillation detection, telemetry mining, antijamming, beam-hopping, and energy management. 

In the absence of solar energy, satellites become dependent on battery energy, which imposes a large load on the battery and reduce the battery lifetime, thereby expanding the costs of space communication networks. To improve the power allocation scheme and battery lifetime in in satellite-to-ground communications using LEO constellations, the authors of \cite{8977503} employed reinforcement learning (RL) to divide the load of overburdened satellites among close satellites with lower load. A similar energy-efficient channel allocation in satellites, implemented using deep RL (DRL), reduced the energy by 67.86\% when compared with that of prior model \cite{9047860}. Conventional modulation/coding modules can be replaced by deep neural networks (DNNs) that can smartly adapt to the environment and hardware \cite{8808168}. Moreover, DNNs are useful for power saving. For example,  DNN-mediated compression prior to data transmission has been suggested to improve latency and save power \cite{kothari2020final}. In fact, AI-chips have remarkably developed in recent years \cite{8808168} and can now running DNNs at low power, which can help to optimize the power costs, and increase the practicality of communication networks, especially for rural areas. DNN autoencoders and a long short-term memory (LSTM) network \cite{ae} have been proposed in \cite{henarejos2019deep} for detecting and classifying interferences, respectively. 




\subsection{AI for Aerial Networks}

AI is promisisng for optimizing UAVs communications (the aerial segment) \cite{bithas2019survey, lahmeri2021artificial}. For example, the authors of \cite{liu2018energy} managed the connectivity and coverage of UAVs using energy-efficient DRL. The suggested technique optimizes the energy efficiency function while considering the fairness, coverage, connectivity, and energy consumption of the communications. Convolutional neural network (CNN), which are routinely used in imaging and image analysis have been suggested for the detection of UAVs, BSs, and ground users. The beamwidth tilt angle can then be adjusted to reduce the interference \cite{8641422}. In \cite{9410462}, artificial neural networks (ANNs) have been suggested for predicting the signal strength of a UAV in smart cities. ANNs are expected to predict the signal strength more accurately than experimental models and are more computationally efficient than theoretical models. The same model could be extended to the rural communication context. In fact, accurately predicting the signal strength and its time availability is mandatory for intelligent decision making in both urban and rural areas scenarios.

UAV communications require stringent energy constraints; imposing these constraints is especially challenging in rural areas. To reduce the energy consumption in drones, the authors of \cite{9498662} optimally routed UAVs across a reduced number of strategic locations. They modeled the problem as a mixed-integer programming problem and evaluated different scenarios using simulations. They showed that clustering is important for reducing the number of strategic locations and saving the energy in drones.

Furthermore, the continuously increasing computation and storage power of newly developed machines has opened opportunities for distributed learning via on-device local data processing. Federated learning (FL) \cite{mcmahan2017communication}, used in applications requiring data security and privacy, learns a shared global model from distributed mobile devices while keeping the training data at each device. FL is a decentralized learning method because the training data are not shared to a central node. In the FL scheme, UAVs or satellites are not obliged to share their data with a central node; instead, they transmit only local updates, which reduce data traffic, thereby increasing the security and reducing the latency of communications. The authors of \cite{zeng2020federated} applied FL to a swarm of UAVs comprising one leading UAV and numerous following UAVs. Each of the followers trains a local FL model using its gathered data and then shares its local model with the leading UAV, which accumulates the received models, generates a global FL model, and shares it with the UAV network. Extending this concept to rural areas, a swarm of UAVs could be randomly positioned (or smartly positioned using AI) and deep learning models could be trained for connectivity purposes or applications such as agriculture or security. For example, the authors of \cite{shiri2020communication} controlled the paths of a massive UAV population from the start point to the end point, considering the possible random wind perturbations, that could cause to disastrous collisions among the UAVs. The model weights of the ANNs among the UAVs were shared via FL.


\subsection{AI for SAGINs}

Further, AI can be used in space-air-ground integration \cite{fourati2021artificial}. However, wireless network optimization using classical techniques may not be practical for SAGINs, as the networks are highly complex and dynamic. Modeling such complex systems is rendered challenging owing to the density, scale, and heterogeneity of the network. In wireless networks, problems are conventionally solved by applying groups of rules obtained from system analysis, which are based on previous experience and domain understanding \cite{8808168}. The authors of \cite{8612450} proposed a CNN is proposed for routing to maximize the general performance of a SAGIN based on the traffic patterns and the persistent buffer size of the satellites. Improving the satellite choice and the UAV position by optimizing the data rate of the source, satellite, UAV, and destination communication is complex because of the high number of moving satellites and the time-varying network structure. To resolve this problem, the authors of \cite{lee2020integrating} jointly optimized the source, satellite, and UAV association and the UAV position using DRL. The mean data rate of the proposed method was 5.74-times higher than that in a direct communication scenario with no UAV and satellite. Moreover, path planning can help the UAVs to optimize their movement when the rate requirements of both aerial and ground equipment are given, which can improve the overall performance of the network \cite{challita2018cellular}. The authors of \cite{8714026}, resolved several emerging issues emerging in communication, including wireless caching, data offloading, and adaptive modulation using a DRL approach. In \cite{9505612}, a SAGINs multidomain network resource was developed using DRL. Moreover, RL can outperform the benchmark scheme in handover management and resource allocation \cite{azari2020machine}. In air-to-ground communications, machine learning algorithms such as random forest and K-nearest neighbors have been suggested for pathloss and delay-spread predictions \cite{yang2019machine}. 

\begin{table*}
\caption{Examples of AI applications.}
\setlength{\tabcolsep}{5pt}
\begin{tabular}{|p{20pt}|p{30pt}|p{80pt}|p{280pt}|p{50pt}|}
\hline
Ref. & Networks & Highlight &Applications & AI Techniques \\
\hline
\hline
\cite{8977503}& Space & Energy Managing &  Efficient power control for LEO satellite-borne batteries & RL\\
\hline
\cite{9047860}& Space& Energy Managing & Energy-efficient channel allocation & RL\\
\hline
 \cite{hu2019deep}& Space & Beam hopping & Beam hopping algorithm in multibeam satellite systems & RL\\
\hline
\cite{8013734}& Space & Prediction & Pathloss forecasting & Decision Tree\\
\hline
\cite{kothari2020final}& Space & Processing & Data compression & DNN \\
\hline
\cite{henarejos2019deep}& Space &Managing Interference &  Interference detection and classification even in high signal to interference ratio & AEs/ LSTM\\
\hline
\cite{7891032}& Space  & Remote Sensing & Land cover and crop types classification & CNN \\
\hline
\cite{9303210}& Space & Optimization & Handoff strategy optimization & CNN \\
\hline
\hline
\cite{liu2018energy}& Aerial & Management & UAV management for coverage & DRL\\
\hline
\cite{wang2021trajectory}& Aerial & Trajectory &  Efficient trajectory optimization of data collection from multiple IoT ground nodes & DRL\\
\hline
\cite{9509955}& Aerial & Positioning &  Cooperative multiagent approach for optimal drone deployment & RL\\
\hline
\cite{8641422}& Aerial  & Detection & UAVs, BSs, and ground users detection & CNN \\
\hline
\cite{9410462}& Aerial  & Prediction & Prediction of UAV signal strengths & ANN\\
\hline
\cite{shiri2020communication}& Aerial & Optimize Learning & Shared learning among UAVs & FL \\
\hline
\hline
\cite{8612450}& SAGINs  & Routing & Routing & CNN \\
\hline
\cite{lee2020integrating}& SAGINs & Optimize Association & Joint optimization of the source, satellite, and UAV association and the UAV position & DRL  \\
\hline
\cite{9505612} & SAGINs & Orchestration & Multidomain network resource orchestration & DRL \\
\hline
\cite{yang2019machine}& SAGINs  & Prediction & Pathloss and delay spread prediction & RF/ KNN \\
\hline
\end{tabular}
\label{tab3}
\end{table*}

\section{Empowering Rural Communities} \label{4}
\subsection{Healthcare}
Connectivity and the use of AI technologies can improve the quality of life of rural dwellers in different ways, for example, improving the health sector in rural areas. In fact, the lack and low quality of healthcare in rural areas, especially in developing countries, contributes largely to the inequality between rural and urban areas. The application of AI technologies in a rural healthcare context can reduce this gap assisting to physicians for better diagnosis and efficient decision making and other well-trained healthcare workers to compensate the absence of physicians \cite{guo2018application}. After investigating AI-powered clinical decision support systems in different rural areas in China and interviewing clinicians on the subject, the authors of \cite{10.1145/3411764.3445432}, summarized the challenges and opportunities of integrating these systems in the rural context. AI (specifically DNNs) and drones have also been proposed for use in screening and detection in rural areas during severe pandemics such as Coronavirus disease, as they can be deployed without direct human involvement \cite{chintanpalliiomt}. 

\subsection{Education}
Connectivity enables communicties in rural areas to benefit from e-learning. Besides allowing children far from schools to access classes from the comfort of their homes, e-learning admits adults to diverse specializations, whereby they can develop skills in different areas to upgrade their occupations and improve their economic situation. Moreover, people in rural areas can access online courses from top universities around the world via massive open online courses and other educational resources online. When people can learn new occupations and improve their knowledge, they can change their jobs or launch a new business, which will benefit the entire community by creating new opportunities for the population and causing economic growth \cite{burlacu2018using}. Moreover, during pandemics, shifting to online learning is necessary to prevent the spread of disease in schools, but this is impossible without connectivity.

\subsection{Agriculture}

Under the constraints of the growing world population and decreasing agricultural lands, the agricultural sector must be supported with more advanced tools and techniques to mitigate several constraints. In fact, smarter farms, healthier crops, and more efficient production are required more urgently than at any previous time. Rural connectivity, IoT \cite{raj2021survey}, and AI \cite{liu2020artificial, karnawat2020future, vincent2019sensors} can all play a major role in achieving this goal. Crop detection \cite{Fourati_2021, crisostomo2020rice},  nutrition control, growth monitoring \cite{hajare2021design}, health diagnosis \cite{nidhis2019cluster}, and disease forecasting \cite{skelsey2021forecasting}, which assits farmers with their decision making, rely on data collected from several sources, such as sensory systems, drone images, weather data and AI predictions. Moreover,  collaboration between farmers, packagers, transporters, distributors, warehouses, and end consumers can be enhanced via connectivity and IoT \cite{raj2021survey}. The authors of \cite{9454888}, presented an application called What2Grow Learning Box that helps rural dwellers to access agricultural contents and learn about agriculture.

\subsection{Disasters}
Fire in forests, mountains, remote and rural areas that happened in the last few years have caused major destruction in various regions of the world. Fire occurrence can be reduced by evaluating and communicating fire threats. IoT approaches enable early fire detection. The authors of \cite{sendra2020lorawan}, presented  a low-cost network that autonomously assesses the level of fire threat and the occurence of forest fires in rural areas. The system comprises several sensors that estimate the temperature, $CO_{2}$, wind speed, and relative humidity of the area. The collected data are stored and processed in a server. Additionally, UAV imaging can recognize fire risks and prevent fires in a timely manner \cite{moumgiakmas2021computer}. 

\subsection{Economy}

\begin{figure}
    \centering
    \includegraphics[scale = 0.82]{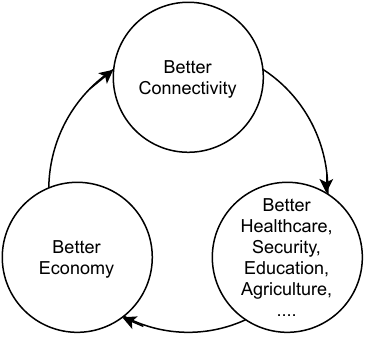}
    \caption{Investing in rural connectivity will improve several sectors including, healthcare, security, education and agriculture, inevitably enhancing economic growth, which attracts more investment in connectivity in rural areas and leads to better connectivity.}
    \label{connect}
\end{figure}

Investing in connectivity in rural areas would markedly improve their healthcare, education, agriculture, and security. Improved healthcare and security systems will ensure a safe and comfortable life for rural dwellers. Furthermore, improved agriculture will enhance the quality and quantity of production, thus increasing the revenue of farmers, while improved education will increase the learning potential of both the young and adult rural dwellers, expanding their knowledge and exposing them to different self developing opportunities, for increasing their skills and thereby changing their occupations or creating new employment avenues. The abovementioned improvements will certainly promote economic growth in rural regions. In turn, a growing economy will the expansion and improvement of connectivity in rural regions, creating a positive feedback effect as summarized in Fig. \ref{connect}.

\section{Discussion}

Solving the challenges of rural connectivity may require several solutions depending different scenarios, including the geography, demography, and more importantly, the resources of the place. When the only constraint is geographical, terrestrial solutions, such as the discussed combined CLOS and NLOS terrestrial network designs, can be used. In more challenging scenarios, for example hard-to-reach or poor rural areas, aerial and satellite networks can play a major role in providing the solutions. Meanwhile, AI can optimize different aspects of communication, including improving the satellite, air, and ground communication networks and their integration to achieve more efficient and practical strategies for rural connectivity.

\section*{Conclusion}
As about half of the world's population is unconnected \cite{gsm2021state}, connectivity in rural areas remains a major obstacle in the communication field. This article briefly overviewed the challenges regarding rural connectivity and discussed a diversity of potential solutions. Connectivity devices can be installed on the ground, flown at low altitudes, or launched in the space. As individual technologies have their advantages and limitations, they have been integrated as SAGINs. Despite their promise for improving rural connectivity, the optimization and management of these technologies remain challenging. AI  appears to be the key to space, air, and ground optimization and integration for rural connectivity. Future work should therefore focus on the use of AI techniques, especially RL and FL for optimal placement and movement of UAVs in rural areas, and SAGIN optimization in different rural contexts.


%





\ifCLASSOPTIONcaptionsoff
  \newpage

\fi



%

\bibliographystyle{IEEEtran}
\bibliography{ref}


%

\end{document}